\documentclass[11pt,twoside]{article}  
\usepackage{asp2006}
\usepackage{adassconf}

\begin{document}   

%
%


%

\title{AWAIC: A WISE Astronomical Image Co-adder}

%
%
%
%
%

\markboth{Masci and Fowler}{A WISE Astronomical Image Co-adder}

%
%
%
%

\author{Frank J. Masci and John W. Fowler}
\affil{Infrared Processing and Analysis Center, Caltech 100-22, 
       Pasadena, CA 91125, USA. Email: fmasci@caltech.edu}

%

\contact{Frank Masci}
\email{fmasci@caltech.edu}

%
%
%

\paindex{Masci, F.}
\aindex{Fowler, J.}     

%



\begin{abstract}          
We describe a new image co-addition tool, AWAIC, to support
the creation of a digital Image Atlas from the multiple frame
exposures acquired with the Wide-field Infrared Survey Explorer (WISE).
AWAIC includes preparatory steps such as frame background matching and
outlier detection using robust frame-stack statistics. Frame co-addition
is based
on using the detector's Point Response Function (PRF) as an interpolation
kernel. This kernel reduces the impact of prior-masked pixels; enables the
creation of an optimal matched filtered product for point source detection;
and most important, it allows for  resolution enhancement (HiRes) to yield a
model of the sky that is consistent with the observations to within
measurement error. The HiRes functionality allows for non-isoplanatic PRFs,
prior noise-variance weighting, uncertainty estimation, and includes 
a ringing-suppression algorithm.
AWAIC also supports the
popular overlap-area weighted interpolation method, and is generic enough
for use on any astronomical image data that supports the FITS and
WCS standards.
\end{abstract}

%
%

\section{Introduction}
The goal of image co-addition is to optimally combine a set 
of (usually dithered) exposures
to create an accurate representation of the sky, given that all
instrumental signatures, glitches, and cosmic-rays have been properly removed.
By ``optimally'', we mean a method which maximizes the signal-to-noise ratio
(SNR) given prior knowledge of the statistical distribution of the
input measurements.

The Wide-field Infrared Survey Explorer (WISE) mission will be
generating over a million exposures (or frames) over the sky.
WISE is a NASA Midex mission
scheduled for launch in late 2009. It will survey the entire sky at 3.3,
4.7, 12, and 23$\mu$m with sensitivities up to three orders of
magnitude beyond those achieved with previous all-sky surveys. For details
on the scientific goals, requirements, instrument and mission design,
see Mainzer et al.\ 2005.
One of the primary products from WISE is a digital Image Atlas that combines
the multiple 8.8 second, $47\arcmin\times47\arcmin$ frame exposures within
predefined tiles over the sky. To support this, we have developed a
suite of software modules collectively referred to as AWAIC for execution
in the automated processing pipeline at the Infrared Processing and
Analysis Center. The modules are written in ANSI-compliant C and wrapped 
into a Perl script, and will be made portable in the near future. 

Here we review AWAIC's co-addition algorithms,
products, and extension to resolution enhancement (HiRes).
It's important to note that HiRes is not in the WISE baseline plan.
It was implemented primarily to support offline research.
The statistical robustness and performance of algorithms will be addressed in
more detail in future papers.

\section{WISE Frame Co-addition Pipeline}
\begin{figure}[t] 
\plotone{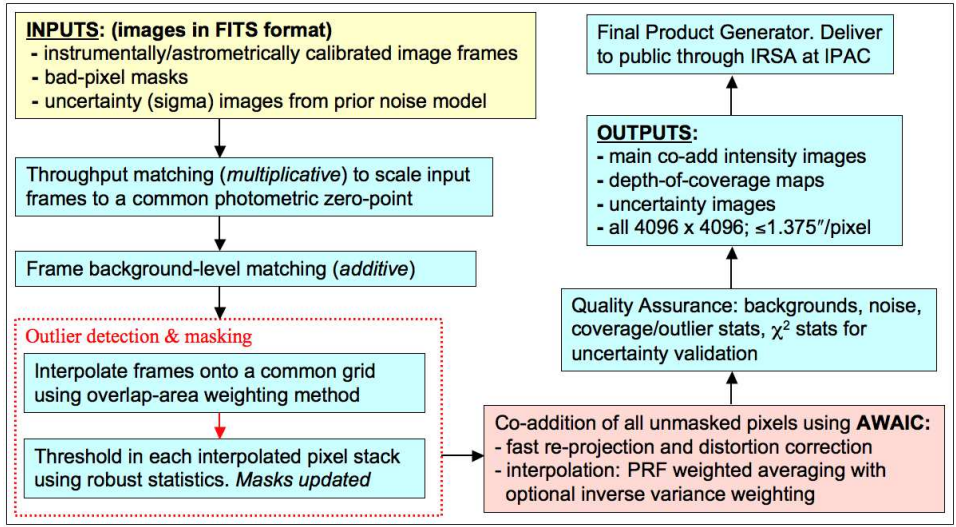}
\caption{Processing steps in WISE frame co-addition pipeline.}\label{fig1}
\end{figure}

Figure~\ref{fig1} gives an overview of the co-addition steps.
It is assumed that the input science
frames have been preprocessed to remove instrumental signatures and
their pointing refined in some WCS using an
astrometric catalog.
Accompanying bad-pixel masks (in 32-bit integer format) and
prior-uncertainty frames are optional.
The frames are assumed to overlap with some predefined footprint (or tile)
on the sky. This also defines the dimensions of the co-add products.
The uncertainty frames store 1-$\sigma$ values for each pixel.
These are expected to be initiated upstream from a noise
model specific to the detector and then propagated and updated as 
the instrumental calibrations are applied. The uncertainties are used
for optional inverse-variance weighting of the input measurements, and 
for computing co-add flux uncertainties.
If bad-pixel masks are specified,
a bit-string template is used to select which conditions
to flag against. The corresponding pixels in the science frames
are then omitted from co-addition. 

The first (optional) step is to scale the frame pixel values to
a common photometric zero-point using calibration zero-point information
in each FITS header. Currently, the software reads a zero-point in
magnitudes stored in the ``MAGZP'' keyword. The common (or target) zero
point is
then written to the FITS headers of the co-add products to enable the 
calibration of photometric measurements.
Frame overlap (or background-level) matching and outlier detection is
then performed. These are described in \S~\ref{prep} Since these initial
steps modify the input frame and mask pixel values, local copies of the
frames and masks are made to avoid overwriting the originals.
After outliers have been flagged in the input masks, the frames are ready
for co-addition. All ``good'' (unmasked) pixels are reprojected and
interpolated onto an upsampled output grid. Details
are described in \S~\ref{coadd} The reprojection uses a fast
input-to-output plane coordinate transformation that implicitly corrects for
focal plane distortion if represented in the input FITS headers.
The Simple Imaging Polynomial (SIP) convention
for distortion is assumed (Shupe et al.\ 2005).

The outputs from AWAIC are the main intensity image, a
depth-of-coverage map, a 1-$\sigma$ uncertainty image
based on input {\it priors}, an image of the outlier locations, and optionally
if the overlap-area interpolation method was used,
an image of the data-derived uncertainty computed
from the standard deviation in each interpolated
pixel stack and appropriately scaled by the depth-of-coverage.
AWAIC also produces a wealth of Quality Assurance (QA) metrics and plots
over pre-specified regions of the co-add footprint. These include
background noise estimates, coverage and outlier statistics, and
metrics to validate co-add flux uncertainties using $\chi^2$ tests.

\section{Preparatory Steps}\label{prep}
\subsection{Background Matching}
Frame exposures taken at different times usually show variations in
background levels due to, for example, instrumental transients,
changing environments, scattered and
stray light. The goal is to obtain seamless (and/or smooth) transitions
between frames near their overlap regions prior to co-addition.
We will want to equalize background levels from frame to frame,
but at the same time preserve natural background variations and
structures as much as possible. We have implemented the 
following simple method:

\begin{enumerate}
\item Fit a robust plane to each frame.
    By ``robust'', we mean immune to the presence of bright
    sources and extended structure. Our goal is to capture the global
    underlying
    background level in a frame. There are of course cases
    where structure
    may span over most of a frame, and hence the background will be
    over-represented. The planar fit is parameterized by $z = f(x,\,y)$,
    where $z$ is the background level, and $x$, $y$ are frame-pixel coordinates.

\item The robust planar fits are subtracted from each respective frame.
    This effectively flattens the frames and places them on
    a zero-baseline.

\item Compute either: (i) the global median $M$ of all frame pixels
    contributing to the co-add footprint, or, (ii) a median plane from
    all the planar fits. The latter attempts to capture any natural
    background gradient over the co-add footprint. 

\item Add this global median $M$ (a constant) to each of the ``zero-level''
    frames (from 2), or, in the case of a median plane, extend it
    over the co-add region and then add it self-consistently to
    each input frame. The frames have now been matched to a common
    background level (or gradient). This will be more or less representative
    of the natural background over the co-add footprint region. 
\end{enumerate}

This method ensures continuity of the background
across the footprint region after co-addition. This not only
improves a co-add's esthetic appearance (by minimizing frame level offsets
between overlaps), but also makes it
self-consistent for scientific purposes.
It's important to note that
instrumental transients or improper instrumental calibration
(e.g., bad flat-fielding) can also manifest as
gradients across frames.
Therefore, one needs to be sure that any retained global gradient is
purely astrophysical.

The above also includes
a method to ameliorate biases from the possible presence of bright extended
structure.
Presence of extended structure (e.g., a galaxy) over a frame 
is first searched for
by thresholding on the ratio of quantile differences in the
pixel distribution, e.g., $Q_d = [q_{0.84}-q_{0.5}]/[q_{0.5}-q_{0.16}]$.
Values of $Q_d\ga 2$ usually indicate a highly skewed
distribution and hence contamination from 
bright extended structure. If detected, a frame
is partitioned into a grid and only those regions with the lowest median
background value are used to perform the robust planar fitting. This method 
still has its limitations, but it extends the robustness of the
algorithm.

\subsection{Outlier Detection and Masking}\label{odet}
The goal of outlier detection is to identify
frame pixel measurements of the same location on the sky which
appear inconsistent with the (bulk) remainder of the sample
at that location. This assumes multiple frame exposures of the
same region of sky are available.
Potential outliers include cosmic rays, latents (image persistence),
instrumental artifacts (including bad pixels), poorly registered frames
from gross pointing errors, supernovae, 
asteroids, and basically anything that has moved or varied 
appreciably with respect to the inertial sky over the observation span 
of a set of overlapping frames.

In summary, the method involves first projecting and interpolating 
each input frame onto a common grid with user-specified pixel scale
optimized for the detector's Point Spread Function (PSF) size. The
interpolation is performed using the overlap-area weighting method 
(analogous to using a top hat kernel).
This accentuates and localizes the outliers 
for optimal detection (e.g., cosmic ray spikes).
When all frames have been interpolated, robust estimates of the
first and second moments are computed
for each interpolated pixel stack $j$. We adopt the sample median ($med$),
and the Median Absolute
Deviation (MAD) as a proxy for the dispersion:
\begin{equation}
\sigma_j = 1.4826\,med\left\{|p_i - med\{p_i\}|\right\},
\end{equation}
where $p_i$ is the value of the $i^{th}$ interpolated pixel within stack $j$. 
The factor of 1.4826 is the
correction necessary for consistency with the 
standard deviation of a Normal distribution 
in the large sample limit.
The MAD estimator is relatively immune to the 
presence of outliers where it exhibits a breakdown point of 50\%,
i.e., more than half the measurements in a sample will need to be declared
outliers before the MAD gives an arbitrarily large error.

The final step involves re-projecting and re-interpolating
each input pixel again, but now testing
each for outlier status against other values in its stack
using the pre-computed robust metrics. A pixel with value $p_i$ 
is declared an outlier if for
given ``upper'' ($u_{thres}$) and ``lower'' ($l_{thres}$) tail thresholds,
either of the following is satisfied:
\begin{eqnarray}\label{out}
p_i &>& med\{p_i\}\,+\,u_{thres}\sigma_j\\\nonumber
p_i &<& med\{p_i\}\,-\,l_{thres}\sigma_j\nonumber
\end{eqnarray}
If declared an outlier, a bit is set in the accompanying frame mask for
use downstream. 
The algorithm also includes
an adaptive thresholding method in that if a pixel is likely to 
contain ``real'' signal (e.g., from a source), the 
upper threshold is automatically 
inflated by a specified amount to avoid (or reduce the incidence of)
outlier flagging at that location.
To distinguish between what's real or not, we generate a
background subtracted {\it median}-SNR co-add
using all the input pixels. 
The background and local noise are computed using spatial median filtering and
quantile differencing: $\sigma\simeq q_{0.5}-q_{0.16}$ respectively.
The idea here is that since these metrics are relatively outlier 
resistant, a large median pixel value in the co-add (or SNR derived therefrom)
is likely to contain signal associated with a source. Therefore,
when flagging outliers using Eq.~\ref{out},
we also threshold on the SNR co-add to determine if
$u_{thres}$ should be inflated.

\begin{figure}[t]
\plotone{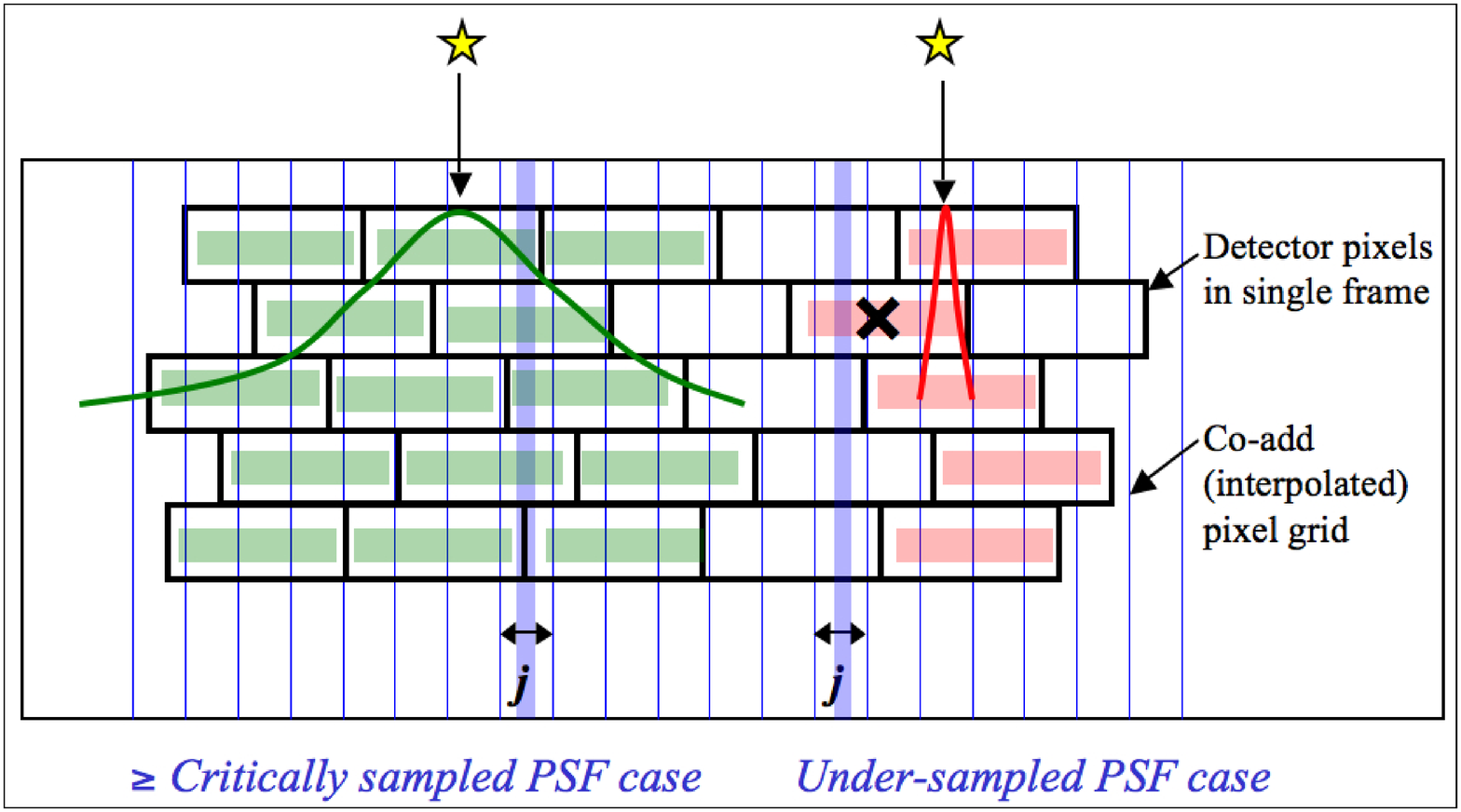}
\caption{One-dimensional schematic of stacking method for detecting outliers
         for well sampled ({\it left}) and under-sampled ({\it right}) cases.
         The input pixel marked ``$\mbox{\boldmath$\times$}$''
         contains signal from a source and is in
         danger of being flagged when outlier detection is performed
         at location $j$ in the output grid.}\label{fig_odet}
\end{figure}

We require typically at least five samples (overlapping pixels) in a stack
for the above method to be reliable. This is because the 
MAD measure for $\sigma$, even though robust, can
itself be noisy when the sample size is small.
Simulations show that for the MAD to achieve the same accuracy as the most 
optimal estimator of
$\sigma$ for a normal distribution (i.e., the sample standard deviation),
the sample size needs be $\simeq2.74\times$ larger.
A noisy $\sigma$ will adversely affect the ability to 
perform reliable outlier detection.
Another requirement to ensure good reliability is to have good sampling 
of the instrumental PSF, i.e., at the Nyquist rate or better.
When well sampled, more detector
pixels in a stack can be made to align within the span of the PSF,
and any pixel variations due to PSF shape are minimized. 
On the other hand, a PSF which is grossly under-sampled can artificially 
increase the scatter in a stack, with the consequence of erroneously flagging
pixels containing true source signal.
Figure~\ref{fig_odet} illustrates these concepts.
The WISE detectors are all slightly better than critically sampled.
Simulations have shown that for depths-of-coverage of eight or more,
(where eight is the median depth for WISE when scanning across the ecliptic),
we expect to detect outliers to completeness and reliability 
levels of $\ga80\%$ for a nominal threshold of $\sim5\sigma$.

\section{Co-addition using PRF Interpolation}\label{coadd}
One of the interpolation methods in AWAIC involves using the detector's
Point Response Function (PRF) as the interpolation kernel.
The PRF is simply the instrumental PSF convolved with the pixel response.
When knowledge of the intra-pixel responsivity is absent, the pixel
response is assumed to be uniform, i.e., a top hat. The PRF is what one 
usually measures off an image using the profiles of point sources. Each
pixel can be thought as collecting light from 
its vicinity with an efficiency described by the PRF. 

The PRF can
be used to estimate the flux at any point in space as follows.
In general, the flux in an output pixel $j$
is estimated by combining the input detector pixel
measurements $D_i$ using PRF-weighted averaging:
\begin{equation}\label{ceq}
f_j\,=\,\frac{\sum_i{(r_{ij}/\sigma_i^2)D_i}}
             {\sum_i{r_{ij}/\sigma_i^2}},
\end{equation}
where $r_{ij}$ is the value of the PRF from input pixel $i$ at the
location of output pixel $j$. The $r_{ij}$ are volume normalized to unity,
i.e., for each $i$, $\sum_j{r_{ij}}=1$. This will ensure flux is conserved.
The inverse-variance weights ($1/\sigma_i^2$) are optional and default to 1.
The $\sigma_i$ can
be fed into AWAIC as 1-$\sigma$ uncertainty frames, e.g., as propagated from
a prior noise model. The sums in Eq.~\ref{ceq} are over
all input pixels in all input frames. With multiple overlapping input 
frames, this will result in a co-add. The 1-$\sigma$ uncertainty in the co-add
pixel flux $f_j$, as derived from Eq.~\ref{ceq} is given by
\begin{equation}\label{seq}
\sigma_j\,=\,\left[\sum_i{w_{ij}^2\,\sigma_i^2}\right]^{1/2},
\end{equation}
where $w_{ij}=(r_{ij}/\sigma^2_i)/\sum_i{r_{ij}/\sigma^2_i}$.
Equation~\ref{seq} assumes the measurement errors
(in the $D_i$) are uncorrelated. Note that this represents the 
co-add flux uncertainty based on priors.
With $N_f$ overlapping input frames and
assuming $\sigma_i=$ constant throughout, it's not difficult to show that
Eq.~\ref{seq} scales as: 
$\sigma_j\simeq\sigma_i/\sqrt{N_fP_j}$, where $P_j=1/\sum_i{r_{ij}^2}$
is a characteristic of the detector's PRF, usually referred to as the
effective number of ``noise pixels''. This scaling also assumes that the
PRF is isoplanatic (has fixed shape over the focal plane) so that  
$P_j=$ constant.
Furthermore, the depth-of-coverage at co-add pixel $j$ is given by the sum of
all overlapping PRF contributions at that location: $N_j=\sum_i{r_{ij}}$.
This effectively indicates how many times a point on the sky
was visited by the PRF of a ``good'' detector pixel $i$, i.e., not rejected
due to prior-masking. If no input pixels were masked, this reduces to 
the number of frame overlaps, $N_f$.

In general, the PRF is usually non-isoplanatic,
especially for large detector arrays. AWAIC allows for a list 
of spatially varying PRFs to be specified, where each PRF corresponds
to some pre-determined region (e.g.,
a partition of a square grid) on the detector focal plane.

Equation~\ref{ceq} can be compared to the popular
pixel overlap-area weighting method, e.g., as
implemented in the \htmladdnormallinkfoot{{\it Montage}}{http://montage.ipac.caltech.edu/} tool.
In fact if the PSF is grossly under-sampled, then the PRF is effectively
a top hat spanning one detector pixel.
The interpolation as described
above then reduces to overlap-area weighted averaging where the
interpolation weights $r_{ij}$ become the input($i$)-to-output($j$) 
pixel overlap areas $a_{ij}$. Incidentally, AWAIC also implements 
the overlap-area weighting method, in case detector PRFs are
not available.

\begin{figure}[t]
\plotone{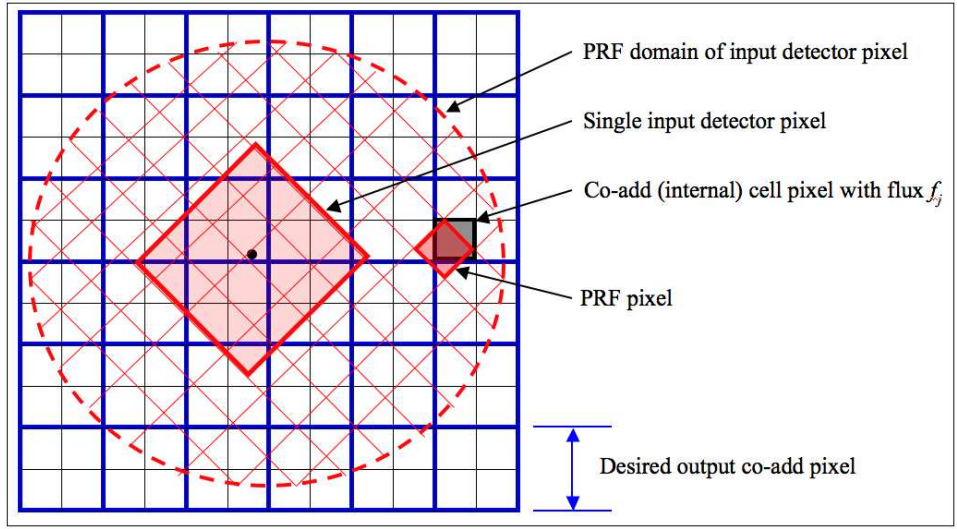}
\caption{Schematic of PRF interpolation for a single input pixel.}\label{fig2}
\end{figure}

Figure~\ref{fig2} shows a schematic of a detector PRF mapped onto the
co-add output grid. The PRF boundary is shown as the dashed circle and
is centered on the detector pixel.
To ensure accurate mapping of PRF pixels and interpolation 
onto the co-add grid, a finer
cell-grid composed of ``pixel cells'' is set up internally.
The cell size
can be selected according to the accuracy to which the PRF
can be positioned.  
The PRF is subject to thermal fluctuations in the optical 
system as well as pointing errors if multiple frames are being combined.
Therefore, it does not make sense to have 
a cell-grid finer than the measured positional accuracy of the PRF.
The PRF pixels are mapped onto the cell-grid frame 
by first projecting the center of the
detector pixel with distortion correction 
if necessary, and then using a local transformation with rotation to
determine the positions of the PRF pixels in the cell-grid. 
The value of a PRF-weighted detector pixel flux $r_{ij}D_i$ in a co-add 
cell pixel $j$ is then computed
using either a nearest-neighbor match, or, the overlap-area weighting method.   
The latter is more accurate but slower.
After all the input pixels with their PRFs have been mapped, the 
internal co-add cells are down-sampled to the
desired output co-add pixel sizes.

There are three advantages to using the PRF as an interpolation kernel.
First, it reduces the impact of masked (missing) input pixels if the 
data are well sampled, even close to Nyquist. This is because the PRF tails
of neighboring ``good'' pixels can overlap and
stretch into the bad pixel locations to effectively give a non-zero
coverage and signal there in the co-add.
Second, Eqs~\ref{ceq} and~\ref{seq} can be used to define 
a linear matched filter optimized for  
point source detection. This effectively represents a cross-covariance
of a point source template (the PRF) with the input data. It leads to
smoothing of the high-frequency noise without affecting the point
source signal sought. In other words, the SNR per pixel
in the co-add is maximized for detecting point source peaks.
The inclusion of inverse-variance weighting further 
ensures that the SNR is maximized since it implies
the co-add fluxes will be
maximum likelihood estimates for normally distributed data.
The creation of co-adds which are also optimized
for source-detection
will benefit projects (e.g., WISE) where a source catalog is also a
release product.
The third advantage is that the PRF allows
for resolution enhancement by
``deconvolving'' its effects
from the input data.

Use of the PRF as an interpolation kernel also has its pitfalls,
at least for the process of co-add generation.
The operation defined by Eq.~\ref{ceq} leads
to a ``smoothing'' of the input data in the co-add grid.
This smoothing is minimized for a top-hat PRF spanning
one detector pixel (equivalent to overlap-area weighting).
This leads to smearing of the input pixel signals and one
consequence is that cosmic rays can masquerade as point sources
(albeit with narrower width) if not properly masked.
For point sources with Gaussian profiles,
their effective width will increase by a factor of $\simeq\sqrt{2}$.
Furthermore, a broad kernel will cause the noise to be
spatially correlated in the co-add, typically on scales (correlation lengths)
approaching the PRF size. Correlations are minimized for
top-hat kernels. Both the effects of flux smearing and correlated
noise must be accounted for in photometric measurements off the co-add,
both in profile fitting and aperture photometry. The compensation
for flux smearing can be handled through an appropriate aperture correction.
Ignorance of correlated noise will cause photometric uncertainties to be
underestimated. Methods on how to account for correlated noise in photometry
will be discussed in a future paper.

\section{Extension to Resolution Enhancement}\label{hires}
We now describe a generic framework for co-addition
with optional resolution enhancement (HiRes). Above we referred to
the concept of combining frames to create a co-add. The HiRes
problem asks the reverse: what model or representation of the sky
propagates through the measurement process to yield the observations 
within measurement error? As a reminder, the measurement process is
a filtering operation performed by the instrument's PRF:
\begin{equation}\label{mproc}
{\rm sky\,(truth)}\,\,\otimes\,\,\underbrace{PSF\otimes\sqcap}_{PRF}\,\,
\otimes\,\,{\rm sampling}\,\,\rightarrow\,\,{\rm measurements.}
\end{equation}
Our goal is to infer a plausible model of the sky or ``truth'' given the
instrumental effects.

\subsection{The Maximum Correlation Method}
The HiRes algorithm in AWAIC is based on the 
Maximum Correlation Method (MCM).
This was originally implemented to boost the scientific return of data
from {\it IRAS} approximately 20 years ago (Aumann et al. 1990;
Fowler \& Aumann 1994), and is still provided as an online service to users.
We have now implemented MCM in a form which is suitable for 
use on any imaging data that are compatible with the FITS and WCS standards,
and the SIP convention for distortion.
The versatility of MCM is that it implicitly generates, as its very 
first step (or first iteration),
a PRF-interpolated co-add as described above. The algorithm is as follows.

\begin{enumerate}

\item First we begin with a flat model image of ones, i.e., a
``maximally correlated'' image:
\begin{equation}\label{mcm1}
f^{n=0}_j\,=\,1\,\,\forall\,\,j,
\end{equation}
where the subscript $j$ refers to a pixel in the upsampled output grid, and
$n$ refers to the iteration number.
This starting image is a first guess at the ``truth'' that we plan to 
reconstruct. Obviously this is a bad approximation, since it represents
what we know without any measurements having been used yet. We could  
instead have used prior information as the starting model if it was available.

\item Next, we use the detector PRF(s) to ``observe'' this model
image, or predict the input detector measurements. Starting with $n=1$,
the predicted flux
in each detector pixel $i$ is obtained by a ``convolution'':
\begin{equation}\label{mcm2}
F^n_i = \sum_j r_{ij}f^{n-1}_j,
\end{equation}
where $r_{ij}$ is the response (PRF value) of pixel $i$ at the location
of output model pixel $j$. Eq.~\ref{mcm2} is a tensor inner
product of the model image with the flipped PRF (see below for why we need
to flip the PRF). It may not be
a true convolution since the kernel $r_{ij}$ may be non-isoplanatic. 

\item Correction factors are computed for each 
detector pixel $i$ by dividing their measured flux, $D_i$, by those
predicted from the model (Eq.~\ref{mcm2}):
\begin{equation}\label{mcm3}
K^n_i = \frac{D_i}{F^n_i}.
\end{equation}

\item For each model pixel $j$, all ``contributing'' correction
factors, i.e., contributed by the overlapping PRFs of all neighboring 
detector pixels $i$ are averaged using response-weighted averaging (with
optional $1/\sigma_i^2$ weighting):
\begin{equation}\label{mcm4}
C^n_j\,=\,\frac{\sum_i{(r_{ij}/\sigma_i^2)K^n_i}}
                      {\sum_i{r_{ij}/\sigma_i^2}},
\end{equation}

\item Finally, the model image pixels are multiplied by their
respective averaged correction factors (Eq.~\ref{mcm4}) to obtain new 
refined estimates of the model fluxes: 
\begin{equation}\label{mcm5}
f^n_j = f^{n-1}_j\,C^n_j.
\end{equation}

\end{enumerate}

If we are after a simple PRF-interpolated co-add, we terminate the process
at step 5. In fact, Eq.~\ref{mcm4} is analogous to the co-addition
equation (Eq.~\ref{ceq}) in that a
starting model image with $f^0_j=1$ implies a correction factor 
$K^1_i\equiv D_i$ since a PRF volume-normalized to unity predicts 
$F^1_i = 1$ (Eq.~\ref{mcm2}). Therefore after the first ($n=1$) 
iteration of MCM, {\it co-add} fluxes will automatically result: 
$f^1_j = f^0_jC^1_j=f_j$.

If we desire resolution enhancement, the above process is iterated, where
the model image from step 5 is used to
re-predict the measurements in step 2. This process of 
iteratively refining the model 
continues until the model reproduces the measurements to within the
noise, i.e., the predictions from Eq.~\ref{mcm2} are consistent with
the measurements $D_i$. If input prior uncertainties ($\sigma_i$) are available,
this convergence can be formally checked
using a global $\chi^2$ test that uses all the input detector pixels:
\begin{equation}\label{mcmchi}
\chi^2_n\,=\,\sum_{i=1}^N{\frac{(D_i - F^n_i)^2}{\sigma_i^2}},
\end{equation}
where we expect $\chi^2_n\simeq N$ (the number of degrees of freedom, $=$ the
number of input pixels).
Alternatively, convergence can also be checked by examining the
correction factors for each detector pixel (Eq.~\ref{mcm3}), where we expect 
$K^n_i\simeq1$ within the noise, or,
via the averaged correction factors (Eq.~\ref{mcm5}), where 
$C^n_j\rightarrow1$ after many iterations.
An image of the latter can be generated by the software at each iteration. 
Iterating much further beyond the initial signs of convergence
has the potential of introducing unnecessary (and usually unaesthetic) detail
in the model. This is important to ensure a parsimonious HiRes solution.

Therefore, it is an algorithmic property of MCM that it only 
modifies (or de-correlates)
a {\it flat} starting model image to the extent necessary to make it 
reproduce the measurements within the noise.
A PRF-interpolated co-add (from the first MCM iteration) will generally not
satisfy the measurements after it is ``convolved'' with the detector PRFs,
i.e., when subject to the measurement process (Eq.~\ref{mproc}).

As a detail, the
input PRFs are first flipped in $x$ and $y$ (or equivalently rotated by $180\deg$)
when HiRes'ing is performed ($n > 1$). This is to conform to the usual 
rules of convolution and
assumes the input PRFs were made by combining images of
point sources observed with the same detector in the same native $x$-$y$ 
pixel frame. For PRF-interpolated co-adds however (that terminate at $n=1$),
the PRFs are not flipped since a cross-covariance with the input data 
is instead needed.
The PRFs here are used as matched filters to generate products 
optimized for point source detection (see \S~\ref{coadd}). 

It is also worth noting that MCM reduces to the classic 
Richardson-Lucy (RL) method if the following are assumed:
(i) the PRF is isoplanatic so that a constant 
kernel allows for Fourier-based 
deconvolution methods to be used; (ii) the inverse-variance weighting 
of measurement correction factors is 
disabled from the PRF-weighted averaging (Eq.~\ref{mcm4}), 
or equivalently if all the input variances $\sigma^2_i$ are assumed equal.
This implies the solution will converge to the 
maximum likelihood estimate for
data that are Poisson distributed. With inverse-variance 
weighting included,
the solution converges to the maximum likelihood estimate 
for Gaussian distributed data.
This is usually always satisfied for astronomical image 
data in the limit of high photon counts;
(iii) there is no explicit testing for global convergence 
at each iteration by checking, for example, that the solution
reproduces the data within measurement error
(Eqs~\ref{mcm2} and~\ref{mcmchi}). This criterion was 
indeed suggested by Lucy (1974), although it is seldom 
used in modern implementations of the RL method.

In the absence of prior information for the starting model, MCM implicitly
gives a solution which is the ``smoothest'' possible, i.e., has maximal
entropy. This should be compared to maximum entropy methods   
(e.g., Cornwell \& Evans 1985)
which attempt to minimize the $\chi^2$ of the differences between the
data and the convolved model, with an additional constraint
imposing smoothness of the solution. MCM requires no explicit 
smoothness constraint. MCM can indeed use a regularizing constraint in the 
form of non-flat starting model, (e.g., an image of the sky from another
detector or wavelength), but this jettisons the idea of an image with 
maximally correlated pixels,
and the refined model image will not be the smoothest possible.
Smoothness is important because it can be used to convey fidelity in a model.
In general, the solution to a
deconvolution problem is not unique, especially in the presence of noise. 
Many models can be made to fit the data, and many methods invoke regularization
techniques in order to select a plausible solution.
A consequence is that some methods give more structure or detail
than necessary to satisfy the data, and there is no guarantee that this
structure is genuine. MCM adopts the Occam's razor approach. 
Given no prior constraints (apart from satisfying the input data), 
MCM will always converge on the simplest solution. This will be the
smoothest possible.

\subsection{The CFV Diagnostic}
\label{cfvd}
A powerful diagnostic from MCM is the Correction Factor Variance (CFV).
This represents the variance about the PRF-weighted average correction
factor (Eq.~\ref{mcm4}) at a location in the output grid for
iteration $n$: $V^n_j=\langle K^2_i\rangle_j - \langle K_i\rangle^2_j$, or
\begin{equation}\label{cfv}
V^n_j\,=\,\sum_i{w_{ij}[K^n_i]^2} - \left[\sum_i{w_{ij}K^n_i}\right]^2,  
\end{equation}
where $w_{ij}=(r_{ij}/\sigma^2_i)/\sum_i{r_{ij}/\sigma^2_i}$,
and the detector-pixel correction factors
$K^n_i$ were defined in Eq.~\ref{mcm3}.
At early iterations, the CFV is generally high everywhere because spatial 
structure has not yet been resolved,
and the model contradicts the
measurements when subject to the measurement process.
If after convergence, all the detector-pixel measurements 
contributing a non-zero response at
some location $j$ agreed 
exactly with their predicted fluxes (Eq.~\ref{mcm2}), 
then all the $K^n_i$ would be $\approx1$ and the CFV ($V^n_j$) at 
that location would be zero.
Areas with a relatively large CFV indicate the presence
of input pixel measurements which do not agree with the majority  
of the other measurements (e.g., outliers). It could also indicate
noisy data, saturated data,
regions where the PRF is not a good match (e.g., erroneously 
broad), or that a field has not yet converged and would benefit from further 
iteration. By thresholding the CFV, one can therefore create a mask for 
a HiRes image to assist in photometry, e.g., to avoid outliers and
unreliable detections from amplified noise fluctuations (see below).

Example CFVs for the M51 galaxy are shown in Figure~\ref{fig4}. The 
corresponding co-add and HiRes'd images appear in Figure~\ref{fig3}.
To illustrate the above concepts, outlying input measurements were not masked 
in the left and middle images of Figure~\ref{fig4}. These refer to
iteration levels $n=1$ and $n=40$ respectively, with the latter
corresponding to convergence.
The CFV image on the far right was created from data with outliers
detected and masked {\it a priori} using the algorithm described 
in \S~\ref{odet} 

Apart from providing a qualitative diagnostic, the CFV can also be used
to compute {\it a posteriori} (data-derived)
uncertainties in the pixel fluxes $f^n_j$ in a HiRes image.
In general, the 1-$\sigma$ uncertainty  
at iteration $n$ can be written in terms of the CFV as:
\begin{equation}\label{sigj}
\sigma^n_j\,=\,c^n f^n_j\sqrt{V^n_j/\sum_i{r_{ij}}},
\end{equation}
where the sum is over the responses from all measurements $i$
at output pixel $j$, i.e., the effective depth-of-coverage.
$c^n$ is a correction factor to account for re-distribution of noise
power across spatial frequencies from one iteration to the next.
At low iterations, power is relatively high at low frequencies, i.e.,
the noise is correlated across pixels. As iterations increase, noise is
de-correlated and power migrates
to high frequencies. The spectrum approaches that of white noise, provided
the input measurement noise was spectrally white.
For $n=1$ (giving a co-add), $c^1\equiv 1/\sqrt{P_j}$, where 
$P_j$ is the effective number of noise pixels defined in \S~\ref{coadd}
With $c^1$ written this way, Eq.~\ref{sigj}
becomes equivalent to the co-add pixel uncertainty defined in Eq.~\ref{seq}.
In general, the $c^n$ at any iteration $n\geq1$ can be approximated 
from the output image products as:
\begin{equation}\label{cn}
c^n\simeq\frac{\sigma_{RMS}[f^n_j]}{\langle\sigma^n_{j}[c^n=1]\rangle},
\end{equation}
where $\sigma_{RMS}$ is the standard-deviation 
(or some robust equivalent) of the
pixel noise fluctuations within a ``source-free'' stationary background region
with $\approx$uniform depth-of-coverage in the $f^n_j$ image.
The denominator is the mean (or median) of 
Eq.~\ref{sigj} with $c^n=1$ in the same region.
At the time of writing, AWAIC only computes an image of
$\sigma^n_{j}[c^n=1]$, since it can be quite subjective on how 
the source-free
stationary background is chosen. If such doesn't exist, background
fitting may be required with $\sigma_{RMS}$ computed from
the fit residuals.
The user can then rescale the $\sigma^n_{j}[c^n=1]$ image
using the estimate of $c^n$ from Eq.~\ref{cn}.
This will give pixel uncertainties which are more or less statistically 
compatible with noise fluctuations in the HiRes'd image.
Pixel SNRs will also be the maximum possible since MCM
would have converged to the maximum likelihood estimate for
data that were Gaussian distributed.
With the correct value of $c^n$, a user then has an estimate of the 
flux uncertainty anywhere in the HiRes'd image, including at the location
of sources. This will allow one to estimate uncertainties in 
source photometry. Noise correlations are also expected to be minimal  
in a converged HiRes image, or negligible if products 
were created with ringing suppression turned on (see below).

\begin{figure}[t]
\plotone{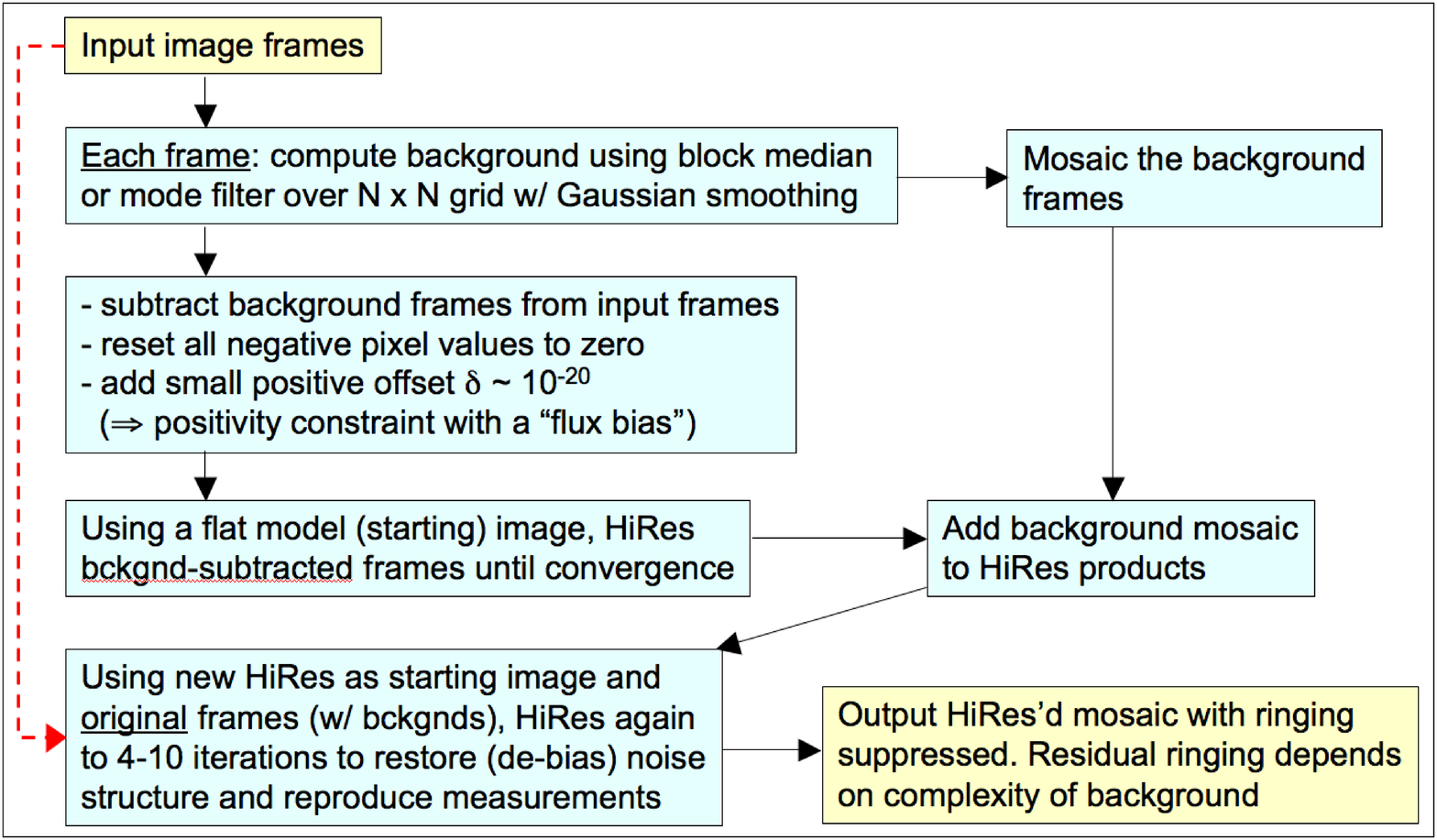}
\caption{Ringing suppression algorithm.}\label{fig_ring}
\end{figure}

\subsection{Ringing Suppression}
Like most deconvolution methods, MCM can lead to ringing artifacts
in the model image. This limits super-resolution, i.e., when attempting 
to go well beyond the diffraction limit of an imaging system.
In general, ringing occurs because the reconstruction process  
tries to make the model image agree with the ``true'' scene
with access to
only the low spatial frequency components comprising the data.
The input data are usually band-limited, and information beyond  
some high spatial frequency cutoff can never be recovered.
The best we can ever reconstruct is a ``low-pass filtered'' version of the 
truth, with the filter determined by the maximum spatial frequency 
the observations provide. This includes the finite sampling by pixels.
A hard high frequency cutoff will lead to sinc-like oscillations
in real image space.
The magnitude of the ringing depends on the strength of a source
relative to the local background intensity level.

It is no accident that a solution with ringing is the smoothest 
(and simplest) solution possible with MCM. Anything smoother 
(with more low frequency power) will not satisfy the measurements
when subject to the measurement process (Eq.~\ref{mproc}).
However, since a large number of less-smooth solutions can
reproduce the observations, those without ringing are
generally more desirable.
Therefore, we relax our request for the smoothest image
and use {\it prior} knowledge that
the background and (desired) source fluxes are 
physically distinct and separable.
There have been numerous approaches that have used this philosophy
(e.g., Lucy 1994).
In brief, the ringing suppression algorithm in AWAIC  
first generates an image of the slowly varying background
for each input frame on some 
specific scale using median filtering;
this is subtracted from the respective input 
frames to create the ``source'' images;
negative noise fluctuations are set to zero, and a tiny positive offset added;
MCM is then run on the background-subtracted images until 
convergence; the background images are combined and then added to the 
HiRes'd source-image product.
This operation enforces a positivity constraint for
reconstruction of the source signals.
It ensures that source flux won't ring against an essentially
zero background level so power can be
forced into high spatial frequencies.
After the background has been added to the HiRes'd source-only product,
MCM is re-executed for several iterations using this as the starting
model image and the original frames as input.
This step re-adjusts the solution and attempts to restore the
intrinsic noise properties of the HiRes process, i.e., what one 
would have obtained
if no background were removed or positivity constraint enforced. It ensures that
photometric uncertainties don't become biased
and the final solution adequately reproduces the measurements 
within the noise. Figure~\ref{fig_ring} gives an overview of the above steps. 

\begin{figure}[t]
\plotone{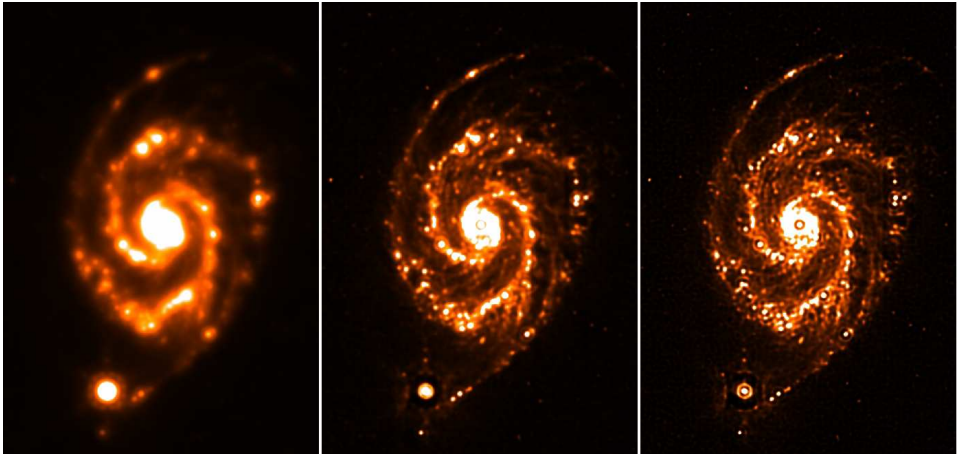}
\caption{M51 from {\it Spitzer}-MIPS 24$\mu$m.
        {\it Left}: co-add (after 1 iteration); {\it Middle}: HiRes
        after 10 iterations; {\it Right}: HiRes
        after 40 iterations.}\label{fig3}
\end{figure}

\subsection{HiRes in Practice}
Like most deconvolution methods, MCM
does not alter the information content of the input image data.
The signal and noise at a given frequency are scaled
approximately together, keeping the SNR fixed.
The process just re-emphasizes different parts of the
frequency spectrum to make images more amenable to a certain kind of
examination, e.g., for detecting previously unresolved objects and
thereby increasing the completeness of surveys.

For optimal HiRes'ing, the input data will have to
adequately sample the instrumental PSF to at least
better than the Nyquist sampling
frequency $2\nu_c$, where $\nu_c$ is the maximum frequency cutoff
inherent in the PSF. For a simple diffraction-limited
system with aperture diameter $D$, $\nu_c\propto D/\lambda$ and corresponds to
the full width at half maximum (FWHM) of an Airy beam.
Even if the detector pixels undersample the PSF (below Nyquist),
redundant coverage with $N$ randomly dithered frames can help recover
the high spatial frequencies, since the average
sampling will scale as $\approx 1/N$ of an input pixel.
The better the sampling, the better the HiRes algorithm is at
improving spatial resolution. For imaging data from the 
{\it Spitzer} IRAC and MIPS detectors
with typically SNR$\ga5/$pixel
and 10 frame overlaps, our HiRes algorithm
reduces the FWHM of the effective PRF to $\simeq0.35\lambda/D$ - a factor
of almost 3 below the diffraction limit. This corresponds to almost an order
of magnitude increase in flux per solid angle for a Gaussian profile.
This enhancement assumes accurate knowledge of the PRF over the focal plane.

\begin{figure}[t]
\plotone{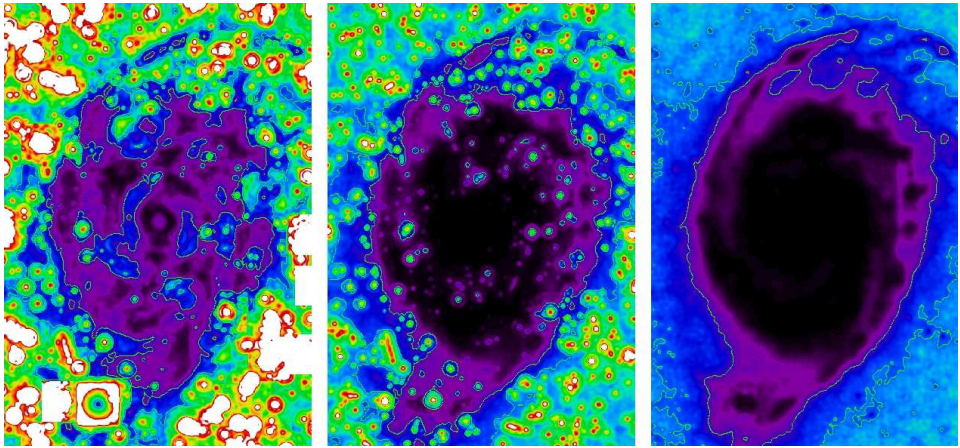}
\caption{Correction Factor Variance (CFV) images for
         M51 whose intensity images were shown in Fig.~\ref{fig3}.
         For a description of the CFV, see \S~\ref{cfvd}
         {\it Left}: CFV after 1 iteration
         with outlying measurements purposefully retained;
         {\it Middle}: CFV after 40 iterations with
         outlying measurements also purposefully retained;
         {\it Right}: CFV after 40 iterations but with outliers
         masked (omitted) prior to HiRes'ing. Darkest regions
         correspond to lowest values of the CFV ($V_j\la0.1$), and
         the brightest to highest values ($V_j\ga100$).}\label{fig4}
\end{figure}

An example output from AWAIC at three MCM iteration levels
is shown in Figure~\ref{fig3}. At high iterations, point source ringing
starts to appear. The ringing around the satellite dwarf galaxy at the   
bottom is aggravated because the core is saturated in the data, and the PRF 
used for HiRes'ing (which is derived from unsaturated data)
is not a good match. ``Flat'' core profiles in the 
data, due to either saturation or improperly corrected non-linearity,
will contain relatively more power in the side-lobes than the actual PRF. 
When this PRF is used for HiRes'ing, these side-lobes will manifest as
ringing artifacts in the HiRes image in order for it to reproduce the
observations on ``convolution'' with the PRF.
Even though the ringing suppression algorithm was turned on in this example,
ringing is still seen around other point sources.
This is because these sources are superimposed on the extended
structure of the galaxy. This structure acts like an elevated background 
against which point sources can ring. 
The ringing suppression algorithm relies on accurate estimation of the
local background, and this can be difficult when complex structure 
is involved, as it is here.

\section{Summary and Future Work}
We have given a broad overview of the algorithms implemented 
in a new generic co-addition/HiRes'ing tool.
The goal is to produce high fidelity science quality products with
uncertainty estimates and metrics for validation thereof.
The HiRes (MCM) algorithm contains considerable improvement over previous 
methods in that it includes {\it a posteriori} 
uncertainty estimation, statistically motivated convergence criteria,
a powerful diagnostic (the CFV) to locate inconsistencies
in the input data and assess the overall quality of HiRes solutions, and the
ability to handle non-isoplanatic PRFs.
Algorithms will be discussed in more detail in future papers.
Future work will explore methods to accelerate 
convergence in MCM,
the ability to handle time-dependent PRFs (e.g., adapted to variable seeing),
and an analysis of the completeness, reliability, and photometric accuracy
of sources detected in HiRes'd images, especially in confused fields.
More examples, analyses, user-interface details, and animations of
MCM can be found at
\htmladdURL{http://web.ipac.caltech.edu/staff/fmasci/home/wise/awaic.html}

\acknowledgements
The authors thank Roc Cutri for guidance and support.
This work was carried out at the California Institute of Technology,
with funding from the National Aeronautics and Space Administration,
under contract to the Jet Propulsion Laboratory.

\end{document}